\documentclass[useAMS,usenatbib,usegraphicx]{mn2e}
\usepackage{amssymb}
\usepackage{graphicx}
\usepackage{subfigure}
\usepackage{longtable}

\newif\ifAMStwofonts

\def\kms{km s$^{-1}$}
\def\hi{H{\sc i}}
\def\hii{H{\sc ii}}
\def\msun{M$_\odot$}

\def\msunyr{M$_\odot$ yr$^{-1}$}

\def\x{$\times$}
\def\deg{$^{\circ}$}
\def\My{M$_{\odot}$ yr$^{-1}$}

\title[The environs of SG\,13]{Multifrequency study of the ring nebula SG\,13}

\author[J. Vasquez, C.E. Cappa and S. Pineault]
{J.Vasquez$^{1,2}$\thanks{Postdoctoral fellow of CONICET, Argentina\
E-mail: jvasquez@fcaglp.unlp.edu.ar}, 
C.E.Cappa$^{1,2}$\thanks{Member of Carrera del Investigador, CONICET, 
Argentina}
and S. Pineault$^{1,3}$\\
\footnotemark[1]
$^{1}$Instituto Argentino de Radioastronom\'{\i}a (CCT-La Plata, CONICET), C.C.5., 1894, Villa Elisa, 
Argentina\\
$^{2}$Facultad de Ciencias Astron\'omicas y Geof\'{\i}sicas, Universidad 
Nacional de La Plata, La Plata, Argentina\\
$^{3}$D\'epartment du physique, de g\'enie physique et d'optique and Centre de recherche en astrophysique du Qu\'ebec (CRAQ),\\ Universit\'e Laval, Qu\'ebec, Canada GIVOA6}

\date{Accepted 2005 June 23, Received 2005 June 22; in original form 2005 March 23}

\pagerange{\pageref{firstpage}--\pageref{lastpage}} 

\pubyear{2008}

\begin{document}

\maketitle

\label{firstpage}

\begin{abstract}

We  investigate the morphology and kinematics of the interstellar medium in the environs of the open cluster Mrk\,50, which includes the Wolf-Rayet star WR\,157 and a number of early B-type stars. The analysis was performed using  radio continuum images at 408 and 1420 MHz, and \hi\ 21cm line data taken from the Canadian Galactic Plane Survey, molecular observations of the $^{12}$CO ($J=1\rightarrow0$) line at 115 GHz from  the Five College Radio Astronomy Observatory  and available mid and far IR observations obtained with the MSX and IRAS satellites, respectively.\

This study allowed identification of the radio continuum and molecular counterpart of the ring nebula SG\,13, while no neutral atomic structure was found to be associated. The nebula is also detected in images in the mid and far infrared, showing the existence of dust well mixed with the ionized gas.
We estimate the main physical parameters of the material linked to the nebula.

The interstellar gas distribution in the environs of Mrk\,50 is compatible with a stellar wind bubble created by the mass loss from WR\,157.

The distribution  of young stellar object (YSO) candidates in the region shows that stellar formation activity may be present in the molecular shell that encircles the ring nebula.

\end{abstract}


\section{Introduction}\label{intro}

Massive stars inject large amounts of energy into the interstellar medium (ISM) through stellar winds, UV radiation, and during their supernova (SN) phase at the end of their lives. Therefore, the morphology, dynamics and energetics of their interstellar environment are strongly modified by the presence of these stars. 

Wolf-Rayet (WR) stars are commonly believed to be evolved O-type stars which have almost reached the end of their nuclear
burning phase (\citealt{con}, \citealt{mae}, \citealt{vh}).
They are characterized by an intense mass flow with mass loss rates of $\sim (1-5)\times 10^{-5}$ \msunyr\ (\citealt{cap04})
and terminal velocities of 1000\,-\,3000 \kms\ (\cite{vh} and references therein) which sweeps up the surrounding interstellar matter. The gas shed by the star and the swept-up interstellar material are piled-up in expanding shells, called {\it interstellar bubbles} (IBs). Stellar wind shocks modify the temperature, presure and density of the surrounding ISM. The strong UV photon flux (with h$\nu\geq$13.6 eV) of these stars ionizes the bubbles, which are detected in the optical and radio ranges as ring nebulae. When the ionization front is trapped in the expanding IBs, these structures have a neutral outer layer which can be detected in the \hi\ 21-cm line emission. The emission distribution of this radio line in the environs of these stars has shown the presence of cavities and expanding shells linked to the stars and their optical IBs (e.g. \citealt{vasq} and references therein). Molecular gas related to IBs has also been found in a number of cases (\citealt{cap01}).\

    Stellar formation may be favoured in the cold and dense outer shells, following the mechanism proposed by  \citealt{Elm77} and \citealt{Elm00}. In their model, massive stars excite an \hii\ region, that expands and sweeps up a shell of shocked cold neutral gas. Eventually, the dense shell of cold neutral swept-up gas formed around these stars fragments and collapses to produce a new generation of stars. Up to now, very few studies dealing with stellar formation in the molecular shells of IBs have been carried out (see for example \citealt{oey}), and this point is still an open question.\    

     In this paper we report the results of a multifrequency study of the gas distribution around WR\,157 and its optical ring nebula based on both infrared and radio data. Stellar formation activity in the region and its relation to the stellar wind bubble are also investigated.\

\section{WR\,157 and its optical ring nebula}\label{cap2}

  Sh2-157 (\citealt{shar}) consists of two different regions: SG\,13 (= Simeiz\,274) at $b>$--0$^{\circ}35\arcmin$ (\citealt{shai}) which has a claw-like appearance, and SG\,14, placed at $b<$--0$^{\circ}35\arcmin$, which is diffuse and irregular in shape. SG\,13 is associated with the open cluster Markarian\,50 (Mrk\,50) (\citealt{turn}, \citealt{lund}, \citealt{smith}) located at ($l,b$) = (111$^{\circ}$21$\arcmin$,--0$^{\circ}$12$\arcmin$), (RA,Dec [J2000]) = (23$^h$12$^m$,+60$^{\circ}$29$\arcmin$). The  brightest member of this cluster is the WR star WR\,157 (= HD\,219460, WN\,5+B1II, ($l,b$) = (111$^{\circ}$19.8$\arcmin$,--0$^{\circ}$13.8$\arcmin$), (RA,Dec [J2000]) = (23$^h$15$^m$ 12.40$^s$, +60$^{\circ}$27$\arcmin$01.8$\arcsec$), \citet{vh}).\

   SG\,13 and Mrk\,50 are located in the Perseus spiral arm. Based on CCD\,$UBV(RI)_C$ photometry, \citealt{bau} derived a photometric distance $d_{\rm{Mrk\,50}}$ = 3.46$\pm$0.35 kpc for the open cluster. \citealt{vh} gives a spectrophotometric distance $d$ = 3.4 kpc for WR\,157, based on its association with the cluster.

\begin{figure}
   \includegraphics[angle=0,width=84mm]{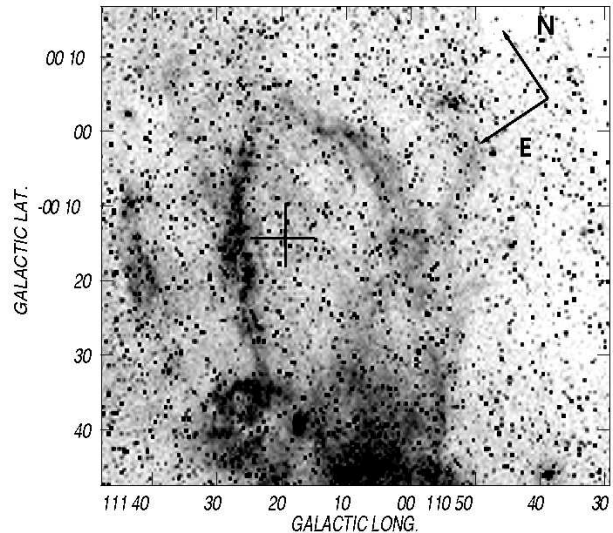}
  \caption{DSS\,R image of SG\,13. The cross marks the location of WR\,157. The grayscale is in arbitrary units. }
  \label{fig1}
\end{figure}

   Fig.~\ref{fig1} displays the DSS\,R image of SG\,13. The cross indicates the position of WR\,157. The ring nebula around Mrk\,50 is easily identified as a claw-like emission region of $\simeq$ 35$\arcmin \times$ 40$\arcmin$ in size, with the open cluster  close to the brightest section of SG\,13. The diffuse emission at  galactic latitudes $b\leq-0^{\circ}35\arcmin$ corresponds to SG\,14.\

 From [OIII]\,5007\AA, [NII]\,6584\AA\ and [SII]\,6717+6731\AA\ observations, \citealt{loz} found that the LSR velocity of SG\,13  is in the range --53 to --33 \kms. This result is consistent with studies of bright regions in SG\,14 based on the H166$\alpha$ RRL (--43 \kms, \citealt{ped}), CO observations (\citealt{blitz}), and H${\alpha}$ data (\citealt{geor75}).

The analytical fit to the circular galactic rotation model by \citealt{bb} predicts that velocities within the range --53 to --33 \kms are located at kinematical distances $d_{\rm k}\simeq$ 3.6 - 5.7 kpc, compatible with the spectrophotometric distance to WR\,157 and Mrk\,50.  A different kinematical distance was published by   \citealt{geor73} and \citealt{gg}, who found  $d$ = 2.5$\pm$0.4 kpc. In what follows, we adopt a distance $d$ = 3.7$\pm$1.2 kpc for SG\,13.\   

\section{Observations}

 The analysis of the gas distribution was performed using observations in the optical, IR and radio ranges. The DSS\,R image of Sh2-157 was obtained from the Skyview web page\footnote{http://skyview.gsfc.nasa.gov/}.\

\subsection{ Radio continuum, HI, and CO data}

 To analyze the ionized and neutral atomic gas distribution, we used \hi\ observations from the Canadian Galactic Plane Survey (CGPS) obtained with the Synthesis Telescope 
of the Dominion Radio Astrophysical Observatory (DRAO) in Canada.
This telescope performed interferometric observations
of the 21-cm \hi\ spectral line, and, simultaneously, continuum
emission in two bands centered at 1420 MHz and 408 MHz.  
The East-West array consists of 7 antennae, 9-m each.
Single-dish data were routinely incorporated into the interferometric images to
 ensure complete coverage of the emission on all angular scales down
 to the resolution limit. Radio continuum observations of the region of SG\,13 have synthesized beams of 3$\farcm$4 $\times$ 3$\farcm$9 and 58$\arcsec\times$ 67$\arcsec$ at 408 and 1420 MHz, respectively. The measured rms image noises are 5 and 1.5 K at 408 and 1420 MHz, respectively. Details about DRAO and the CGPS can be found in \citealt{Land00} and \citealt{tay03}.\ 

    To investigate the neutral hydrogen distribution, we extracted a data cube centered at ($l,b,v$) = (111$^{\circ}$8$\arcmin$ , --0$^{\circ}$23$\arcmin$ , --40.2 \kms) from the CGPS. The \hi\  data have a synthesized beam of 1${\farcm}$13$\times$0${\farcm}$98, a rms noise of 3 K in bright temperature ($T_B$), and a velocity resolution of 1.3 \kms\ with a channel separation of 0.824 \kms. The \hi\ images were convolved to a 2${\arcmin}\times$ 2${\arcmin}$ beam size to facilitate the identification of structures. The observed velocities cover the range --60 to 163 \kms.\

  The $^{12}$CO ($J=1\rightarrow0$) line data at 115 GHz were obtained using the radiotelescope of the Five College Radio Astronomy Observatory (FCRAO), in USA. The angular resolution is approximately 46$\arcsec$. Details about the CO survey are summarized by \citealt{rid}.\

\begin{figure}
   \includegraphics[angle=0,width=84mm]{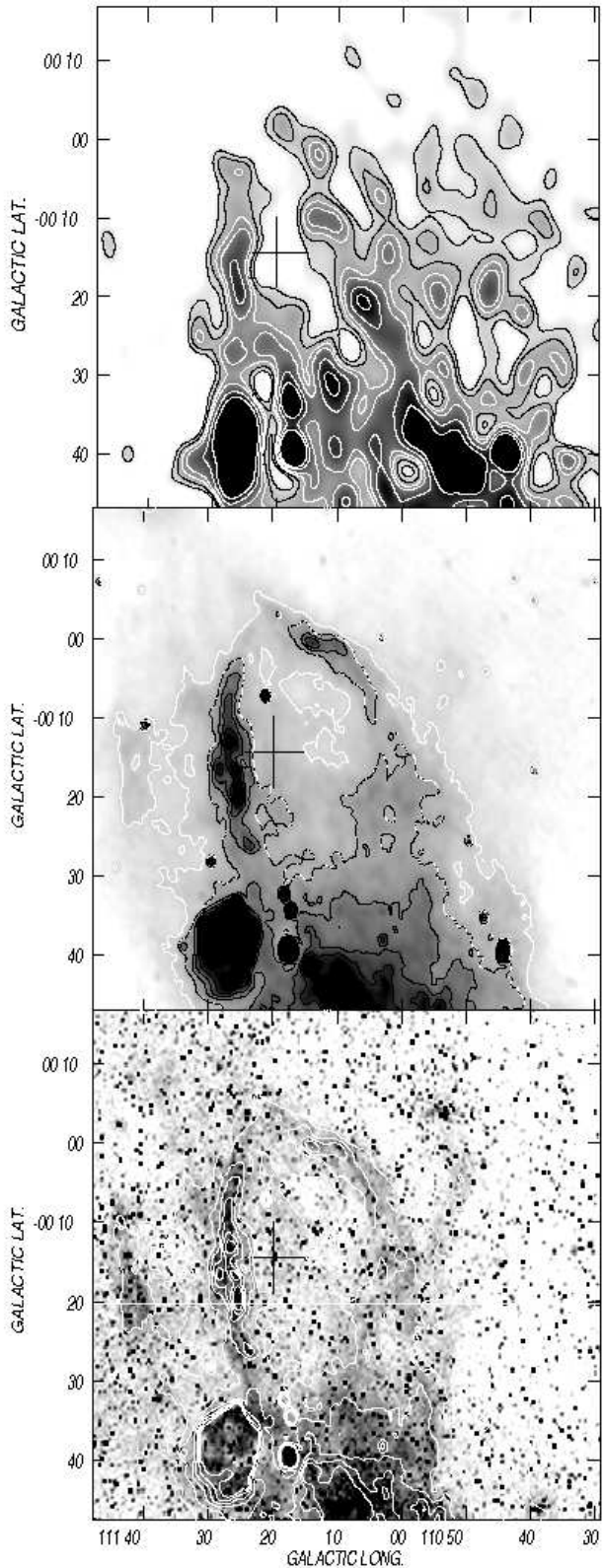}
  \caption{{\it Top panel}: Radio continuum image at 408 MHz. The grayscale in brighness temperature $T_b$ corresponds to 110 to 150 K. The contour lines are 115, 120, 125 130, 140, and 150 K. {\it Middle panel}: Radio continuum image at 1420 MHz. The grayscale in $T_b$ corresponds to 7 to 13 K. The contour lines are 8, 9, 10, 11, 12, and 13 K. The cross marks the position of WR\,157. {\it Bottom panel}: Overlay of the optical emission and the 1420 MHz contour lines.}
  \label{4081420}
\end{figure}

\subsection{IRAS and MSX data}

 Infrared images at different wavelengths were used to analyze the dust 
distribution in the region. High resolution IR images (HIRES) at 60 and 100 $\mu$m, were taken from IPAC
\footnote{IPAC is founded by NASA as part of the IRAS extended 
mission under contract to Jet Propulsion Laboratory (JPL)
and California Institute of Technology (Caltech).} with angular resolutions of $\simeq$ 1$\arcmin$ and 2$\arcmin$.\

 The Spatial Infrared Telescope (SPIRIT\,III) on board the Midcourse Space Experiment (MSX) satellite surveyed the entire Galactic Plane in four mid IR bands centered at 8.28, 12.13, 14.65, and 21.3 $\mu$m (bands A, C, D and E, respectively). We retrieved images in the four bands with 18$\farcs$4 in angular resolution from IPAC. We converted the flux densities to Janskys by using the conversion factor of radiance to flux density given in the MSX Image Server at IPAC (\citealt{egan}).\

To investigate the presence of pre-main sequence stars towards the region under study we extracted IR point sources from the MSX, 2MASS, and IRAS point source catalogues.\

\section{Gas and dust distribution}

\subsection{Radio continuum emission}

   The radio continuum images at 408 and 1420 MHz are shown in the top and middle panels of Fig.~\ref{4081420}. The radio emission distribution at 1420 MHz correlates strikingly well with the optical emission (see the bottom panel of Fig.~\ref{4081420}). The brightest optical emission region coincides with the strongest radio emission region in SG\,13 at 1420 MHz. In addition to the ionized branches present at $l\simeq$ 111$^{\circ}$5$\arcmin$ and 111$^{\circ}$25$\arcmin$, the image at 1420 MHz reveals weak radio continuum emission at ($l,b$) $\simeq$(111$^{\circ}$35$\arcmin$,--0$^{\circ}$15$\arcmin$), coincident with a region of diffuse optical emission. Previous high resolution radio continuum studies of Sh2-157 were centered on SG\,14 and did not include the northern SG\,13 area (\cite{isr}).\

   The ring nebula around WR\,157 can also be identified in the radio continuum surveys at 4850 MHz (Condon et al. 1991) and 2700 MHz (\cite{fu90}), taken with the Effelsberg 100-m telescope. These images are not included in this paper.\

  Clearly, the image at 408 MHz is contaminated with reduction artifacts from the strong source Cas\,A, located 2$^{\circ}$ away from Sh2-157. The sources detected at ($l,b$)=(111$^{\circ}$10\arcmin,--0$^{\circ}$30\arcmin) and ($l,b$)=(111$^{\circ}$5\arcmin,--0$^{\circ}$22\arcmin\ ), which are not evident at 1420 MHz, coincide with the X-ray source X 1WGA\,J23138+6024 and the radio source NVSS\,J231310+601236, respectively. Both are listed in the NED\footnote{NASA/IPAC Extragalactic Database} database as extragalactic radio sources. These sources were not included in the flux density estimate.\

   Derived flux densities (S$_{\nu}$) at 408, 1420 and 2700 MHz are 1.8$\pm$0.9, 3.4$\pm$0.9 and 3.5$\pm$1.0 Jy, respectively. The uncertainty in flux density arises in both the rms noise of the images and in the estimate of the background emission. 

  Uncertainties in the flux densities are too large to allow a meaningful
   determination of the spectral index, however the flux densities are
   consistent with a thermal origin.\

   The sources centered at ($l,b$) = (111$^{\circ}$26$\arcmin$,--0$^{\circ}$40$\arcmin$), ($l,b$) = (111$^{\circ}$18$\arcmin$,--0$^{\circ}$40$\arcmin$) and ($l,b$) = (111$^{\circ}$8$\arcmin$,--0$^{\circ}$46$\arcmin$) in the image at 1420 MHz were named G\,111.4-0.7, S\,157\,A and G\,111.2-0.8 by \citealt{isr}, and correspond to SG\,14.

%\begin{figure}
%\begin{center}
%  \includegraphics[angle=0,width=84mm]{Vasquez3.eps}
%  \caption{log S$_{\nu}$ {\it vs}  log $\nu$ plot for SG\,13. Flux densities are indicated with their errors bars.} 
%\label{flujo}
%\end{center}
% \end{figure}

\begin{figure}
   \includegraphics[angle=0,width=84mm]{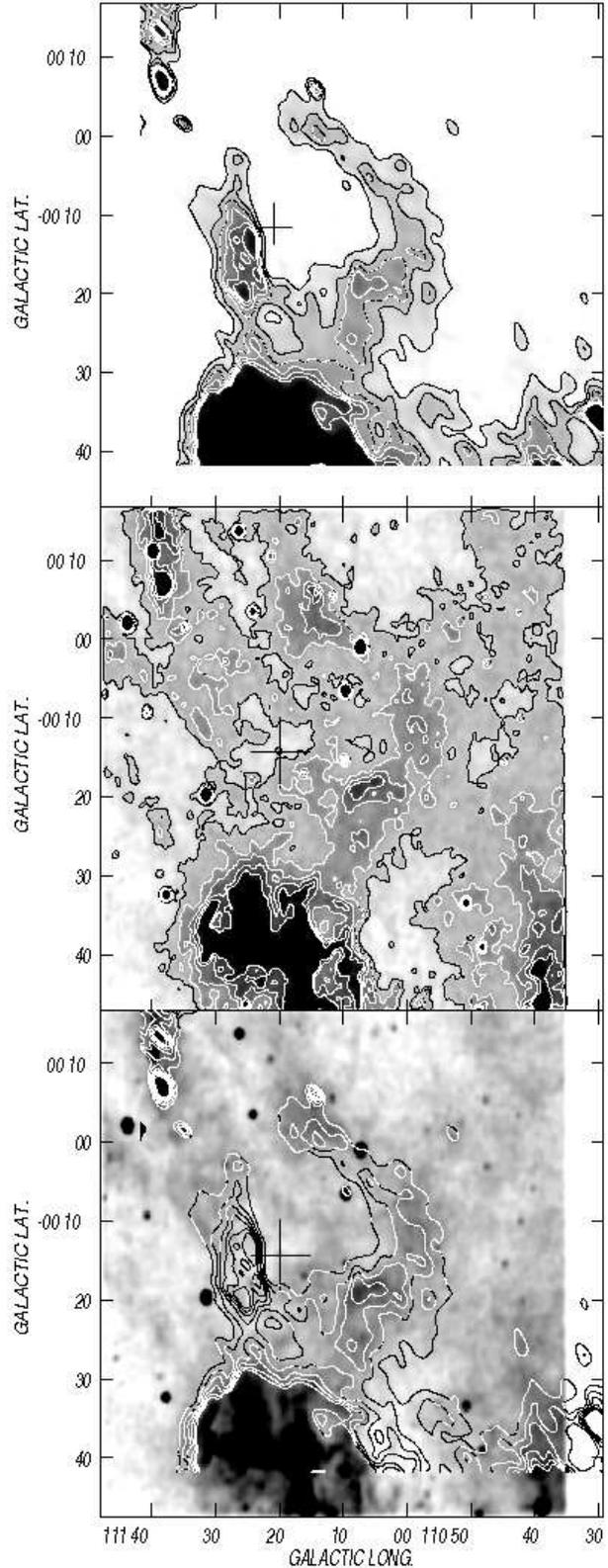}
  \caption{{\it Top panel}: IRAS image at 60 $\mu$m. The grayscale corresponds to 65--140 MJy/sr. The contour lines are 70, 80, 90, 100, 110, and 120 MJy/sr. {\it Middle panel}: MSX band-A at 8.28 $\mu$m. The grayscale is 10.5 (7$\sigma$) to 40 MJy/sr and the contour lines are 16, 21, 26, 31 and 36 MJy/sr. {\it Bottom panel}: Overlay of the 60 $\mu$m (in contour lines) and the 8.3$\mu$m (grayscale) emissions.}
  \label{ir}
\end{figure}

\subsection{Infrared emission}

The {\it IRAS} (HIRES) image at 60 $\mu$m around the open cluster is displayed in the top panel of Figure~\ref{ir}. The middle panel displays the MSX band A emission centered at 8.28 $\mu$m. The 7.6 and 8.6 $\mu$m features of the policyclic aromatic hydrocarbons (PAHs) strongly contribute to the emission within this band. The bottom panel of Fig.~\ref{ir} shows an overlay of the 60 $\mu$m (contours) and 8.28 $\mu$m (grayscale) emission distributions.\ 

     The emission distribution in the far IR presents the same morphology as in the optical and radio continuum bands. The IR emission at 100 $\mu$m, not shown here, displays the same distribution as the 60 $\mu$m image. The emission in the far IR originates in thermal emission from dust. The strong correlation between the far IR and the radio continuum emission suggests that large grains and ionized gas are well mixed.\

      Strong correlation between the emissions at 60 and 8.28 $\mu$m is clear at ($l,b$) $\simeq$ (111$^{\circ}$12$\arcmin$, +0$^{\circ}$2$\arcmin$) and ($l,b$) $\simeq$ (111$^{\circ}$5$\arcmin$ , --0$^{\circ}$20$\arcmin$). The 8.28 $\mu$m emission distribution coincides with the emission at 60 $\mu$m mainly at the right branch at $l\leq$ 111$^{\circ}$20$\arcmin$. A different behaviour is observed towards ($l,b$) = (111$^{\circ}$25$\arcmin$,--0$^{\circ}$15$\arcmin$), where the bright region at 60$\mu$m coincides with a low emission region at 8.28 $\mu$m. Very probably, the dust grains responsible for the emission at 8.28 $\mu$m have been destroyed in this part of the nebula, which is closest to the WR star.\

   The strongest IR emission region present at $b<$--0$^{\circ}$30$\arcmin$ corresponds to SG\,14.\

\subsection{The emission from the molecular gas}

\begin{figure*}
\includegraphics[angle=0,width=170mm]{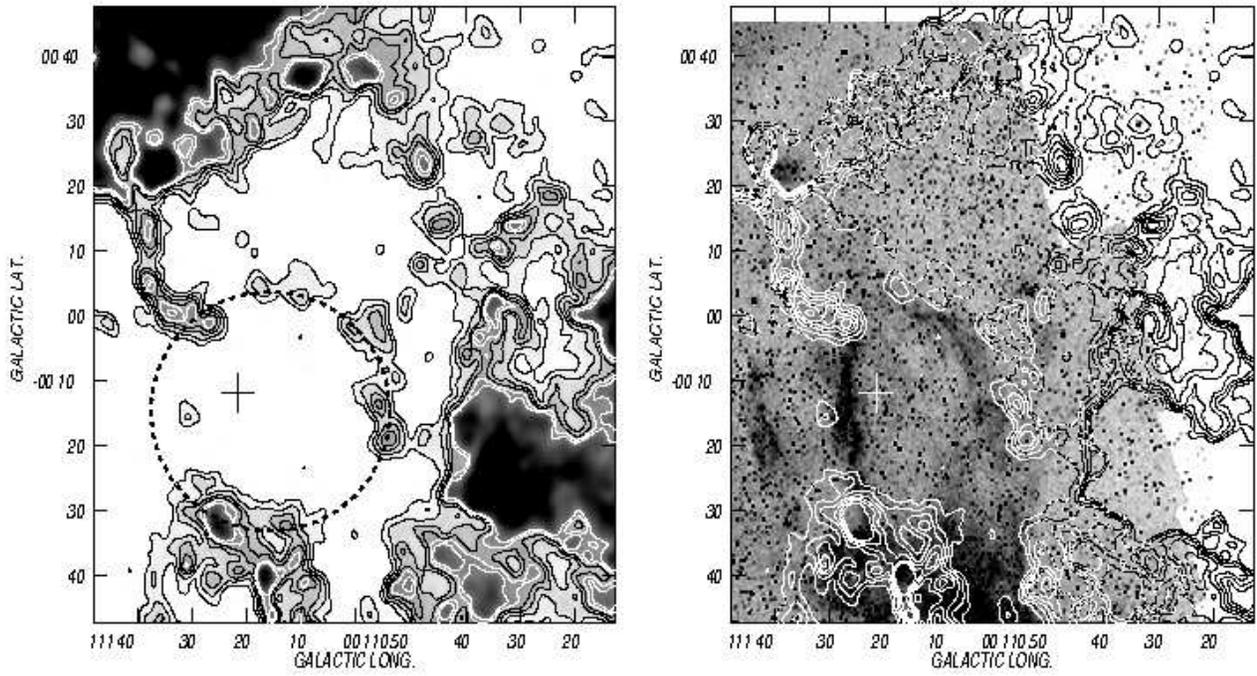}
% \vspace{1cm}
    \caption{{\it Left panel}: Integrated CO emission distribution within the velocity range --58.4 to --43.5 \kms. The grayscale corresponds to 1.5-50 K \kms. The contour lines are 2.5, 6, 11, 16, 21, and 26 K \kms. The dashed circle marks the clumpy CO structure. {\it Right panel}: Overlay of the DSS\,R image in grayscale and the CO emission distribution in contour lines.}
  \label{co}
\end{figure*}

 The CO(1-0) emission distribution of a larger region is shown in Fig.~\ref{co}. The left panel shows the result of integrating the emission within the velocity  range --58.4 to --43.5 \kms\ in grayscale and contour lines, while the right panel shows an overlay of the contour lines of the left panel and the optical emission. A number of CO cloudlets borders the section of the ring nebula toward lower galactic longitudes. The brightest optical region, which runs parallel to $l= 111^{\circ}25\arcmin$, appears almost free of molecular material, since only the end point near $b\simeq -0^{\circ}5\arcmin$ matches a strong CO cloudlet. The CO cloud near ($l,b$) = ($111^{\circ}25\arcmin,-0^{\circ}32\arcmin$) coincides with the low latitude edge of the ring nebula and separates it from SG\,14. Note that the maximun in the CO emission corresponds to an optically faint region. The fragmentary molecular emission delineates a roughly circular structure surrounding the optical ring nebula, as would be expected where an ionized region is surrounded by neutral gas. This is schematically indicated by a dashed circle in the left panel of Fig. \ref{co}.\

   The velocities of the CO gas that surrounds the optical nebula span the interval --55 to --43 \kms. CO associated with the region at ($l,b$) = (111$^{\circ}$30${\arcmin}$,--0$^{\circ}$15${\arcmin}$) was detected between --43 and --47 \kms, while the CO cloud at ($l,b$) = (111$^{\circ}$25${\arcmin}$,--0$^{\circ}$35${\arcmin}$) is present at $\simeq$ --50 \kms, and the chain of CO cloudlets spans the velocity interval --55 to --50 \kms. The region at ($l,b$)=(111$^{\circ}$35${\arcmin}$,+0$^{\circ}$5${\arcmin}$), apparently unrelated to SG\,13, is detected in the velocity range --58 to --48 \kms. The positional correlation between the optical and CO rim structures and the agreement between the velocities of these cloudlets with the corresponding velocity of the ionized gas (see Sect. \ref{cap2}) suggest that the molecular feature is the molecular counterpart of SG\,13. The clumpy CO morphology suggests that most of the CO gas in the region has been dissociated by the strong UV photon flux of the massive stars in Mrk\,50. Considering the whole molecular velocity range, the expansion velocity of the CO ring is 8$\pm$1  \kms. 

  A large and intense patch of CO emission is present in the northern section of the image. With the exception of the \hii\ region Sh2-159, located at ($l,b$) = (111$^{\circ}$36${\arcmin}$,+0$^{\circ}$22${\arcmin}$) (\citealt{isr}), this molecular emission shows little correlation with the optical emission. The strong CO emission projected onto Sh2-159 is very probably associated with the \hii\ region, because of the similar CO and H$\alpha$ velocities (\citealt{blitz}).\

   A comparison of the IR emission distributions at 8.28 and 60 $\mu$m (Fig.~\ref{ir}) with the molecular one (Fig.~\ref{co}) reveals a clear morphological correlation. The CO clouds encircle the IR emission associated with SG\,13.\

   Adopting a systemic CO velocity of --48 \kms\ for the molecular gas related to SG\,13, the circular galactic rotation model by \citealt{bb} predicts a kinematical distance $d_{CO}$ = 5.0$\pm$0.8 kpc. The uncertainty was derived adopting a velocity dispersion of 6 \kms\ for the interstellar gas. The kinematical distance is compatible with the spectrophotometric distance of Mrk\,50 and WR\,157 (see Sect. 2).\

\subsection{The emission of the neutral atomic gas}

\begin{figure*}
\includegraphics[angle=0,width=84mm]{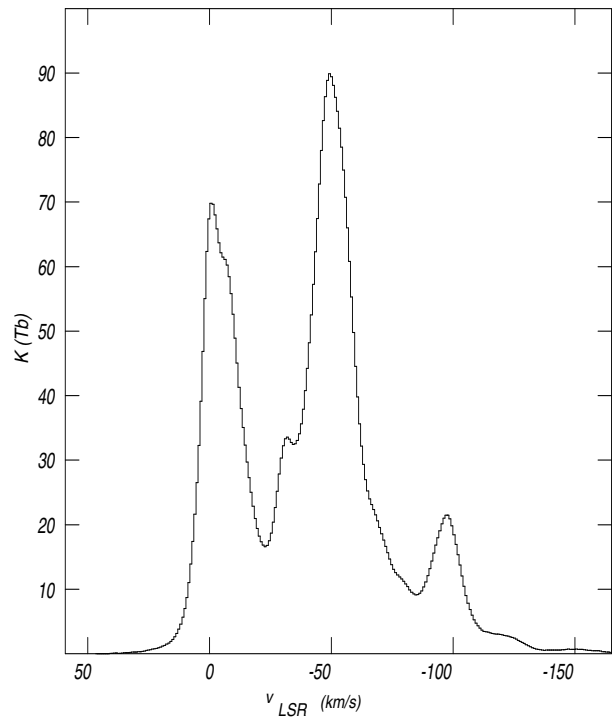}
% \vspace{1cm}
    \caption{Average \hi\ spectrum towards SG\,13}
  \label{corte}
\end{figure*}

  Figure \ref{corte} exhibits the average \hi\ emission distribution along the line of sight  to SG\,13. The profile was obtained by averaging the \hi\ emission within a box of 1$^{\circ}$30$\arcmin$\x 1$^{\circ}$45$\arcmin$ enclosing the \hii\ region. The more intense peaks are present at 0 and --50 \kms\ with brightness temperature  $T_b$ $\sim$ 70 and 90 K, respectively, enclosing a central region with a brightness temperature of about  $\sim$ 20 K. A less intense peak is  centered at $\sim$ --100 \kms\ with  $T_b \simeq$ 20 K. An analytical fit to the circular galactic rotation model by Brand and Blitz (1993) predicts that material with velocities of --100 and --50  \kms\ should be located at $\sim$12 and  $\sim$5 kpc, corresponding to the Cygnus and Perseus arms, respectively, while gas at about 0 \kms\ corresponds to the Orion arm (local gas).\

\begin{figure*}
  \includegraphics[angle=0,width=170mm]{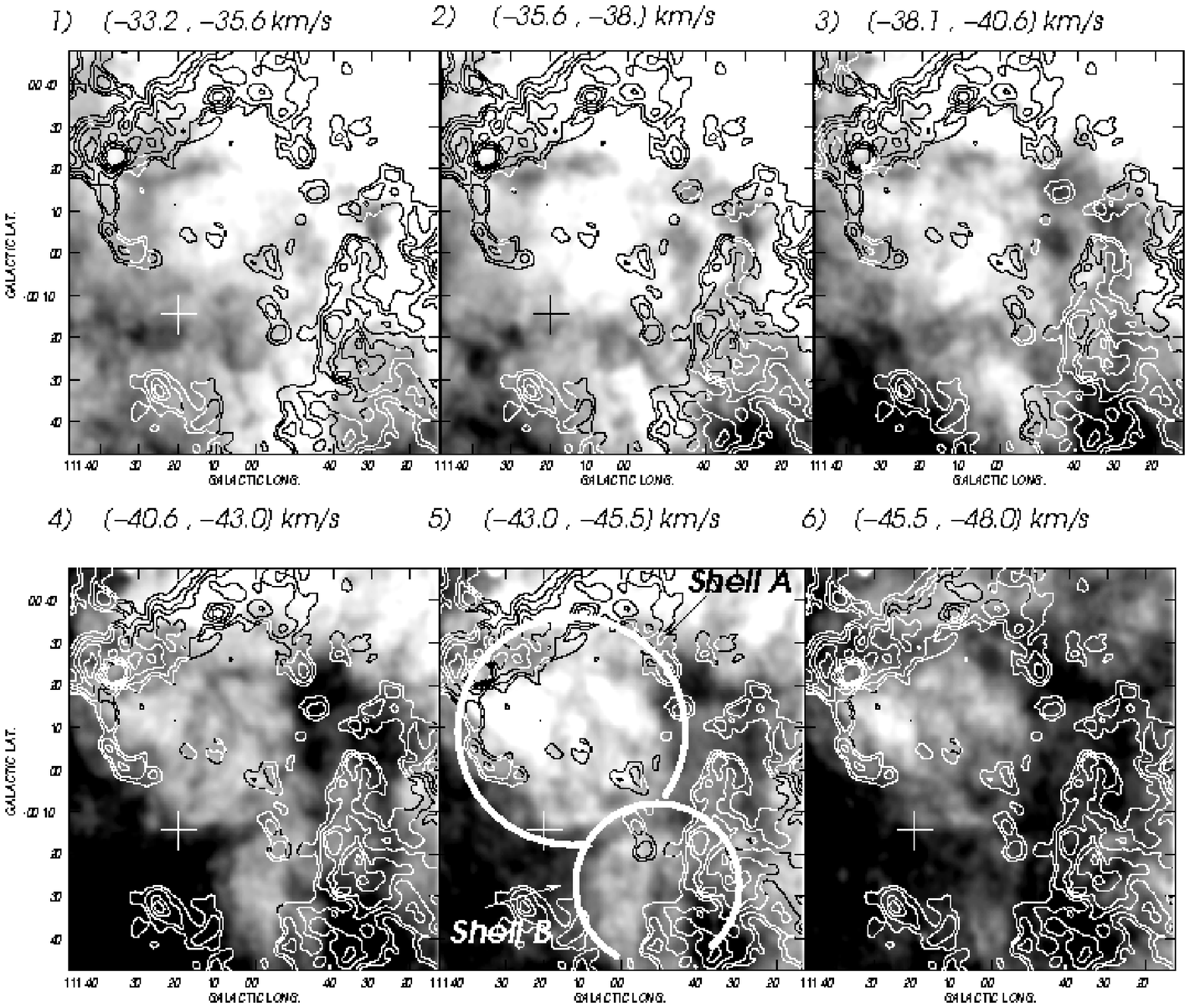}
 \caption{Superposition of the \hi\ emission distribution in the line of sight to Mrk\,50 (grayscale) with the CO integrated emission (contours). Each \hi\ map shows the mean brightness temperature within an interval of $\sim$ 2.5 \kms. The grayscale in $T_b$ of the \hi\ emission is 40 to 100 K for maps (1), (2), (3), and (4); and 70 to 135 K for maps (5) and (6). The cross marks the position of WR\,157. The white circles delineate Shells A and B detected in \hi.}
 \label{hi-co}
\end{figure*}

Figure~\ref{hi-co} displays the superposition of the CO emission distribution shown in Fig.~\ref{co} (in contour lines) and the \hi\ brightness temperature $T_b$ distribution within the velocity range --48.0 to --33.2 \kms\ in steps of 2.5 \kms. The analysis of the \hi\ emission distribution within this velocity range allows identification of two structures at velocities compatible with the radial velocities of the ionized and molecular material linked to SG\,13. The larger and weaker structure, of about $\sim$30$\arcmin$ in radius, is centered at ($l,b$)=(111$^{\circ}$7$\arcmin$,+0$^{\circ}$8$\arcmin$) (hereafter named Shell A), while the smaller and brighter one, of about 20$\arcmin$ in radius, is centered at  ($l,b$)=(110$^{\circ}$55$\arcmin$,\hbox{--0$^{\circ}$35$\arcmin$}) (hereafter Shell B). Both shells are shown in Fig.~\ref{hi-co} as white circles.\

    Shell A is clearly identified at $\simeq$ --44 \kms. The section of this \hi\ structure at $b\simeq$ +0$^{\circ}$20$\arcmin$ is associated with intense CO emission detected at these galactic latitudes.\

  Shell B surrounds part of the \hii\ region SG\,14. The section of Shell B at $l$ = 110$^{\circ}$47$\arcmin$ coincides with relatively strong CO emission, while the border at $l$ = 111$^{\circ}$20$\arcmin$ correlates with faint CO emission linked to SG\,14. However, a morphological connection with the IR and radio continuum counterparts of SG\,13 is not observed. No CO is detected at the interface between the two \hi\ shells.\

\begin{figure*}
\includegraphics[angle=0,width=170mm]{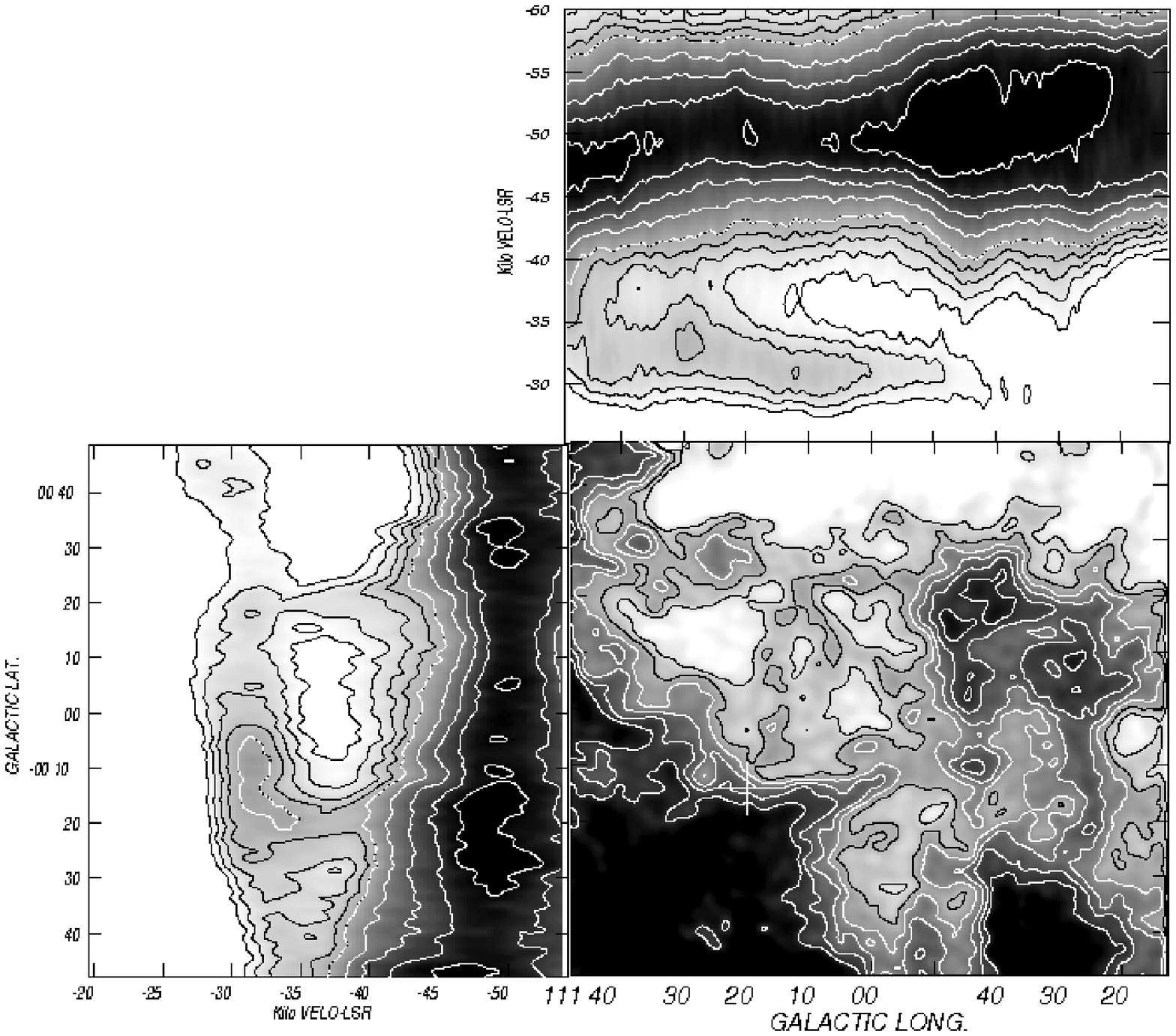}
% \vspace{1cm}
    \caption{{\it Bottom right panel}: \hi\ integrated emission in the velocity interval from -44.0 to -41.5 \kms, corresponding to the systemic velocity of Shells A and B. The grayscale is 80-140 K and the contour lines are 70, 80, 90, 100, 110, 120, 130, and 140 K. {\it Upper panel}: ($v,l$) image for $b$ = 0.0$^{\circ}$ showing brightness temperature $T_b$. Grayscale: 30 to 90 K. Contours: 30, 35, 40, 45, 50, 60, 70, 80, and 90 K. {\it Bottom left panel}: ($v,b$) image for $l$ = 111$^{\circ}$9.6${\arcmin}$ showing the same grayscale and contours as the upper panel.}
  \label{HI}
\end{figure*}

   The lower right panel of Figure~\ref{HI} displays the \hi\ integrated emission in the velocity interval from --44.0 to --41.5 \kms, corresponding to the systemic velocity of Shell A. The upper and left panels of Fig.~\ref{HI} show the ($v,l$) image corresponding to $b$ = 0.0$^{\circ}$ and the ($v,b$) image for $l$ = 111$^{\circ}$10$\arcmin$, respectively. These images allow us to estimate the extension in velocity of Shell A and are useful to derive its expansion velocity. Shell A can be detected within the velocity interval from  --32 to --50 \kms, with an expansion velocity of 13$\pm$2 \kms. Shell B can also be identified on the left panel of Fig.~\ref{HI}, although it is less conspicuous that Shell A. Following Brand \& Blitz (1993) we determine a kinematical distance for Shell A of 4.6$\pm$0.8 kpc.\

The molecular gas related to the section of SG\,13 at ($l,b$) = (111$^{\circ}$10$\arcmin$, 0$^{\circ}$0$\arcmin$) appears projected close to the center of Shell A. The distribution of the molecular material related to SG\,13 correlates neither with Shell A nor with Shell B. The analysis of the \hi\ emission distribution spanning the range \hbox{--48.5} to --51.8 \kms does not show an obvious \hi\ counterpart to the ionized and molecular gas related to SG\,13. \

\begin{figure*}
\includegraphics[angle=0,width=84mm]{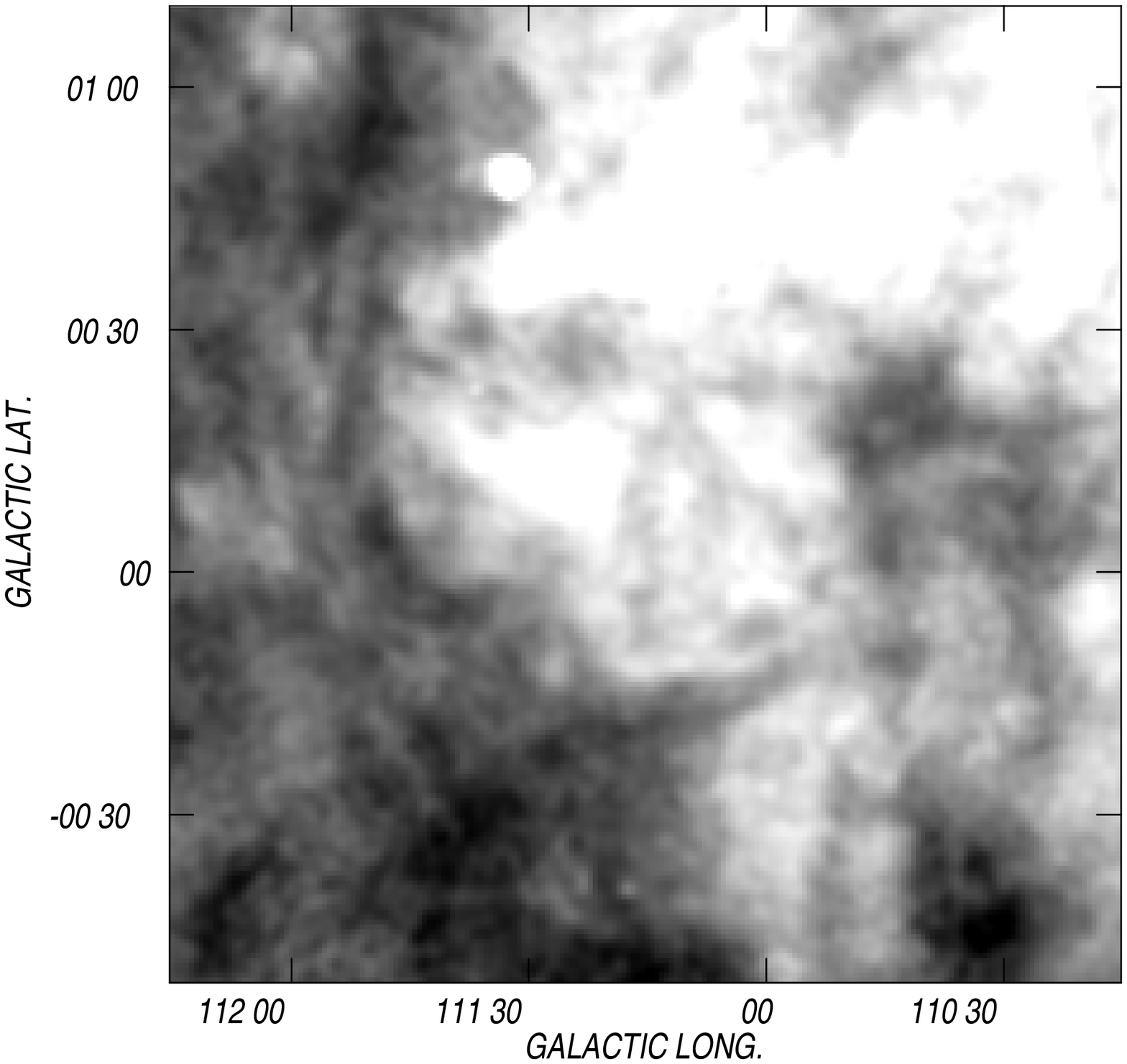}
% \vspace{1cm}
    \caption{General \hi\ environment around SG\,13. The grayscale is the same as in Fig.~\ref{hi-co}.}
  \label{gran-hi}
\end{figure*}

   In order to obtain a better view of the general HI environment of the
two structures, we present in Fig.~\ref{gran-hi} a slightly larger field of view than the one used for the previous \hi\ images. The figure is the average of three CGPS channels between --41.86 and --44.34 \kms, smoothed to a resolution of {\bf $2\arcmin$}. The figure shows very clearly the shell-like structure of feature A.\

\section{Star formation}

    In this section, we consider the possibility that star formation is going on in the expanding shell pushed by the stellar winds of the massive stars in Mrk\,50, particularly by  WR\,157, following the ``collect and collapse'' model by \citealt{Elm77}. We look for the presence of YSO candidates that are projected onto the molecular envelope of SG\,13 in the MSX, IRAS, and 2MASS point source catalogues. The searched box area was 1\deg \x 1\deg\ centered on the position of the open cluster.\

\begin{figure}
\centering
\includegraphics[angle=0,width=84mm]{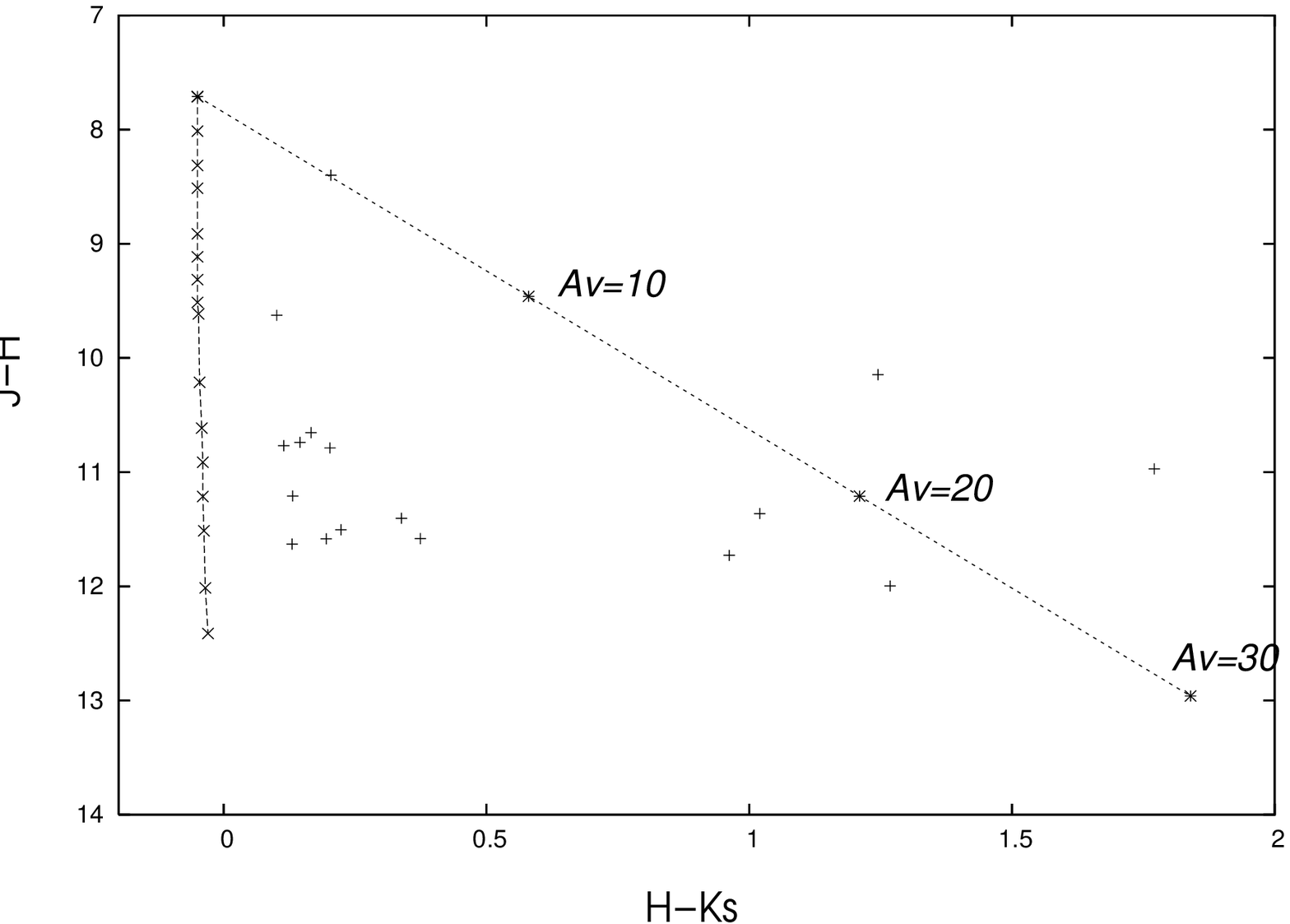}
\caption{CM diagram of 2MASS point sources with IR excess in direction to SG\,13.}
\label{fig3}
\end{figure}

    We follow the \citealt{co05} criteria in looking for YSO candidates in the 2MASS point sources catalogue. These criteria discriminate between giant and main sequence stars with or without reddening, and sources with IR excess. The last ones are the most important sources for our purpose since their IR fluxes reveal the presence of circumstellar IR emission. According to these criteria, we found around 20000 2MASS sources, of which only 19 present IR excess. Fig.~\ref{fig3} displays the colour-magnitude (CM) diagram of the 19 sources with IR excess assuming a distance of 3.7$\pm$1.2 kpc. The ZAMS from O3 to B5 type stars is indicated at the left of the diagram. Half of the IR excess sources present visual absorption larger than 10 mag. Table~\ref{tabla2} summarizes the main parameters of these sources, i.e. galactic coordinates, designation, and $J$, $H$, and $Ks$ magnitudes.\

 Following the criteria given by \citealt{ju}, we have also found that several IRAS point sources projected onto this region are protostellar candidates. The data for these sources are compiled in Table~\ref{tabla2}, which shows the ($l,b$) coordinates, the IRAS name, the fluxes at 12, 25, 60 and 100 $\mu$m of each source, and the FIR luminosity according to Chan \& Fich (1995). We found that 9 out of the 17 IRAS sources are YSO candidates.

\citealt{lums} derived several criteria to help identify massive young stellar objets (MYSOs) in the MSX point-source catalogue. The main idea is to discriminate among sources with IR excess originating in dust envelopes around young and evolved stars, and \hii\ regions. From mid-IR colour-colour diagrams, they found that MYSOs have IR fluxes with ratio $F_{21}/F_8>$2 and  $F_{14}/F_{12}>$1, where $F_8$, $F_{12}$, $F_{14}$ and $F_{21}$ are the fluxes at 8.28, 12.13, 14.65 and 21.30 $\mu$m. For compact \hii\ regions, these flux ratios are $F_{21}/F_8>$ 2 and $F_{14}/F_{12}<$ 1. Evolved stars have $F_{21}/F_8<$ 2. On this basis, no YSO candidates are found inside the searching box area.

\begin{figure*}
\begin{center}
\includegraphics[angle=0,width=170mm]{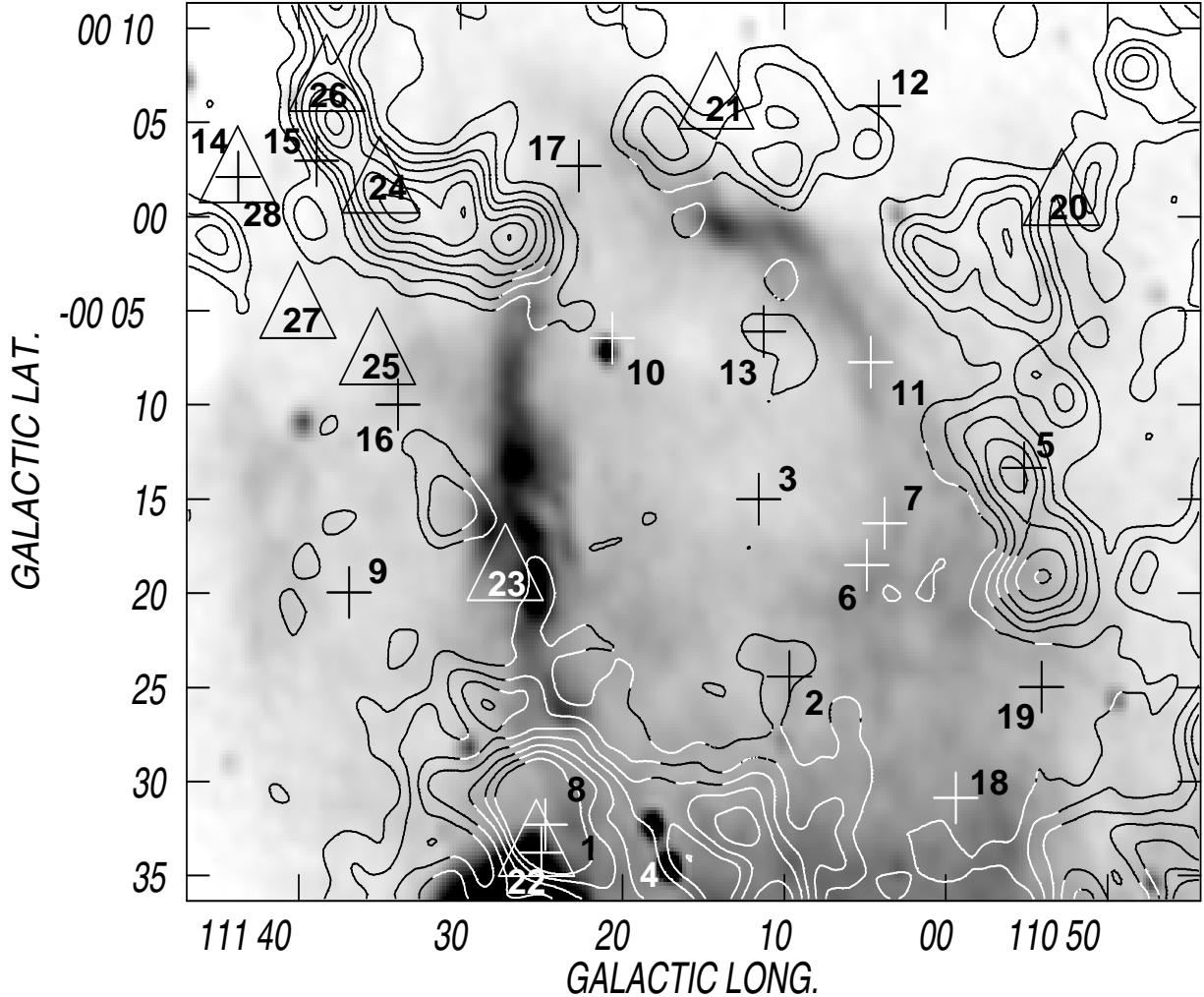}
\caption{Spatial distribution of YSO cantidates in direction to SG\,13 over the image at 1420 MHz in grayscale from 7 to 13 K, and the $^{12}$CO image with the following contours: 1.5, 5, 10, 20, 30, and 40 K \kms. The 2MASS and IRAS sources are indicated by crosses and triangles, respectively. } 
\label{fig4}
\end{center}
\end{figure*}

     In Fig.~\ref{fig4} we display the spatial distribution of the YSO candidates of Table~\ref{tabla2} projected onto the 1420 MHz image in grayscale and the $^{12}$CO(1-0) emission distribution in contours. The crosses correspond to objects with IR excess from the 2MASS catalogue and the triangles to IRAS protostellar candidates.\

     IRAS sources 20, 21, 24, and 26 are projected onto CO cloudlets. Sources 25 and 27 appear projected over regions lacking molecular emission, while source 23 coincides with an ionized filament. On the other hand, 2MASS sources 6, 7, 11, 18, and 19 are projected over radio continuum filaments, while sources 2, 5, 12, 13, and 15 coincide with CO clouds. Sources 9, 16, and 17 are projected onto a diffuse ionized region. Sources 1, 4, 8, and 22 are related to the \hii\ region SG\,14. Finally, sources 14 and 28 are located far away from the molecular emission linked to SG\,13.

     We conclude that many YSO candidates are projected onto the interstellar bubble envelope. Besides, we believe that about half of the 2MASS sources having IR excess with $A_v> 10 $ mag may evolve towards massive stars. Their strong visual absorption and their location in the CM diagram favour this suggestion. The existence of  a relatively large number of 2MASS and IRAS  YSO candidates reveals the presence of young stellar objets in different evolutionary phases. The spatial distribution of the YSO candidates is suggestive of the action of the ``collect and collapse'' process described by \citealt{Elm77}.

\begin{table*}
\begin{minipage}{170mm}
\caption{YSOs candidates in direction to SG\,13}
\label{tabla2}
 \begin{tabular}{ccccccccc}
\hline\hline
\multicolumn{7}{c}{2\,MASS sources} \\

 $\#$ & $l$[$\circ$]          &     $b$[$\circ$]  & Designation & $J$ & $H$ & $K_s$ \\
  &   &   &               & mag & mag & mag \\
\hline\hline

1 & 111.41 & -0.56 & 23164713+6010319 & 15.132 & 13.265 & 11.997 & \\
2 & 111.16 & -0.40 & 23142490+6013421 & 11.428 & 11.34 & 11.209  &\\
3 & 111.19 & -0.24 & 23141131+6023109 & 8.756 & 8.603 & 8.399  &\\
4 & 111.29 & -0.68 & 23161127+6001148 & 14.002 & 12.06 & 10.642 & \\
5 & 110.91 & -0.22 & 23120304+6018385 & 13.062 & 11.392 & 10.147  &\\
6 & 111.08 & -0.30 & 23133130+6017273 & 13.473 & 12.383 & 11.363 & \\
7 & 111.06 & -0.27 & 23131651+6019056 & 11.844 & 11.76 & 11.63 & \\
8 & 111.41 & -0.53 & 23164107+6011498 & 13.713 & 12.691 & 11.729 & \\
9 & 111.61 & -0.33 & 23173713+6027406 & 9.727 & 9.727 & 9.626  &\\
10 & 111.34 & -0.10 & 23145423+603425 & 12.391 & 11.956 & 11.582 & \\
11 & 111.07 & -0.12 & 23125722+6027223 & 11.979 & 11.728 & 11.505 & \\
12 & 111.06 & 0.097 & 23121250+6039495 & 10.977 & 10.884 & 10.739 & \\
13 & 111.18 & -0.10 & 23134236+6031201 & 11.932 & 11.779 & 11.584  &\\
14 & 111.72 & 0.03 & 23172558+6050436 & 7.461 & 5.98 & 4.826  & \\
15 & 111.64 & 0.04 & 23164590+6049490 & 15.606 & 12.742 & 10.971  &\\
16 & 111.56 & -0.16 & 23164539+6035553 & 10.929 & 10.82 & 10.654 & \\
17 & 111.37 & 0.04 & 23144274+6043428 & 11.194 & 10.99 & 10.788 & \\
%18 & 111.31 & 0.22 & 23134205+6052217 & 10.998 & 10.937 & 10.795 & \\
18 & 110.99 & -0.51 & 23132696+6003570 & 10.932 & 10.883 & 10.769 & \\
19 & 110.90 & -0.41 & 23122993+6007256 & 12.191 & 11.742 & 11.404 & \\
%21 & 111.16 & -0.70 & 23151790+5957040 & 13.024 & 12.235 & 11.567 & \\

\hline
\hline
\multicolumn{8}{c}{IRAS sources }\\

 &             &         &   &     \multicolumn{4}{c}{Fluxes} & $L_{\rm FIR}$\\         
  &      &             &                  &  12$\mu$m [Jy] &    25$\mu$m[Jy] &   60$\mu$m[Jy] &  100$\mu$m[Jy]& [Jy] \\
\hline

20  &  110.88 &  0.01 & 23088+6014  & 0.38  & 0.41       & 6.38  &  28.79& 157.92 \\
21  &  111.23 &  0.10 & 23113+6027 &  0.54 &  0.83 & 8.64 &   40.40 & 220.55 \\ 
22  &  111.42 & -0.56 &  23146+5954  & 7.17 &   15.       &   275.        &  671. & 4324.25 \\   
23 &  111.45 &-0.31  &  23141+6008  & 0.46      &   0.21      &  8.22  &  38.79 & 210.13\\
24 &  111.58 & 0.02  & 23141+6030  & 0.71   & 1.05  & 9.60  &  44.90 &245.51\\
25 &  111.58 &-0.12  &  23146+6022 & 0.81  & 0.97 & 5.88  &  32.70 &174.00\\
26 &  111.63 &  0.11 & 23143+6036 & 0.93  &  1.30 &  28.39  &  116.  &  648.31 \\   
27 &   111.66 & -0.08 & 23151+6026 &  0.95 &  0.52 & 4.40 &   23. &  123.88 \\    
28 &   111.73 & 0.03  & 23152+6034  & 22.20  &  25.10 &  147.        &  227. & 1800.00   \\

\hline
\end{tabular}
\end{minipage}
\end{table*}

\section{Discussion}

\subsection{Main physical parameters of the ring nebula}

  Table~\ref{tabla1} shows the more relevant parameters of the ionized and neutral structures linked to Mrk\,50.\ 

   The electron density $n_e$, the ionized mass $M_{\rm ion}$, the emission measure $EM$, and the observational excitation parameter $U_{\rm rad}$ were determined using the classical expressions by \citealt{mh} from the image at 1420 MHz. We assumed that SG\,13 is a spherical nebula with constant density. We adopted an electron temperature $T_e$ = 10$^4$ K, and assumed singly ionized He (the derived mass $M_{\rm ion}$ was multiplied by 1.27 to take this fact into account). The estimated rms electron density and the ionized mass, corresponding to a filling factor $f=1$, are listed in Table~\ref{tabla1}. Electron densities for $f=0.1-0.2$ are also included in Table~\ref{tabla1}. These filling factors were derived by taking into account that SG\,13 is a ring of $\sim$ 12$\arcmin$ in radius and 5$\arcmin$ in thickness, and that only 10$\%$-30$\%$ of its surface is covered by plasma.\

     Based on [SII] emission lines, \citealt{loz} determined electron densities $n_e'=100-150$ cm$^{-3}$ for the optical nebula, higher than the values derived from radio continuum observations. Following Israel (1978), a different estimate for the filling factor can be obtained as $f={(ne/ne')}^{0.5}$ = (2-7)$\times$10$^{-3}$. Electron densities derived from optical line ratios are higher than the ones obtained from radio continuum observations. The former corresponds to regions with high emissivity and high electron densities, while the latter have a large contribution from low electron density regions. Very probably, the real $f$ is in between these two values.\

 $U_{\rm rad}$ is related to the number of UV photons used to ionize the gas and is obtained as $U_{\rm rad}=R {n_e}^{2/3}$, where $R$ is the radius of the ionized gas. $U_*$ is linked to the number of UV photons emitted by the massive stars in the open cluster, which can be obtained from atmosphere models. To estimate $U_*$ we took into account only WR\,157 (WN5 star), since the other stars in Mrk\,50 have spectral types later than B3 (\citealt{bau}). Following \citealt{smi}, $U_*$ $\simeq$ 128 pc cm$^{-2}$. From Table~\ref{tabla1}, $U_{\rm rad}<U_*$, indicating that the UV photons emitted by the WR star are enough to ionize the gas. A large number of UV photons probably escape to the ISM through the clumpy molecular envelope around SG\,13 and/or are absorbed by the associated dust, warming and destroying it.\

          Bearing in mind that the interstellar dust radiates in the far IR, we derived the dust mass $M_{\rm d}$ associated with SG\,13 and the dust colour temperature $T_{\rm d}$ from the emission at 60 and 100 $\mu$m (Draine $\&$ Lee 1984). The absorption coefficient $\chi_{\nu}$ of the dust can be written as

\begin{equation}
\chi_{\nu} = 4 \left(\frac{\nu}{3\times 10^{12} \rm Hz}\right)^{n} \rm{kg^{-1} m^2,} 
\end{equation}

\noindent{where $n=1-1.5$ is the dust spectral index. $T_{\rm d}$ can be obtained as:}

\begin{equation}
T_d (K) = \frac{95.94}{\ln \left[ (1.67)^{3+n} \frac{S_{100}}{S_{60}}\right]},
\end{equation}

  $S_{100}$ and $S_{60}$ are the IR fluxes at 100 and 60 $\mu$m. For $n=1-1.5, $ $T_d\simeq$ 28 to 31 K. Following \citealt{hi83} the dust mass is

\begin{equation}
M_{\rm d,\nu} = \frac{S_{\nu} d^2}{\chi_{\nu} B_{\nu}(T_{d})},  
\end{equation} 

\noindent{where $B_{\nu}(T_{{\rm d}})$ is the blackbody function}. The flux densities at 60 and 100 $\mu$m, along with the derived dust mass are listed in Table~\ref{tabla1}.\
 
   The associated molecular mass of each cloudlet $M_{{\rm H_2}}$, can be obtained as:

\begin{equation}
{M_{\rm H_2}} = \mu\ m_H\ d^2\ \Omega\ {N_{H_2}}.  
\end{equation} 

\noindent{We adopted $\mu=2.76$ for the molecular weight (assuming solar abundances), $m_H$ is the atomic hydrogen mass, $\Omega$ is the solid angle of the molecular cloudlet and ${N_{\rm H_2}}$ is the H$_2$ column density, which is obtained as} 

\begin{equation}
{N_{\rm H_2}} = X \times I_{CO}, 
\end{equation}  

\noindent{where X = 1.9$\times 10^{20}$ $\rm \frac{mol\ cm^{-2}}{K\ kms^{-1}}$ (\citealt{gl90},~\citealt{di95}). $I_{CO}$ is the integrated emission of the CO line. The mean column density and the molecular mass are listed in Table~\ref{tabla1}.}\

     The gas-to-dust ratio derived taking into account the ionized and molecular masses listed in Table~\ref{tabla1} is $\sim$ 20, lower than the typical value of $\sim$ 100 generally accepted for \hii\ regions.

\begin{table}

\begin{minipage}{240mm}
\caption{Physical parameters of SG\,13's counterparts}
\label{tabla1}
 \begin{tabular}{l c c}
\hline

Distance adopted  &   &    3.7$\pm$1.2 kpc \\    
\hline   

      \multicolumn{3}{c}{{\it Radio continuum}}   \\
\hline
%S$_{4850}$(Jy)    &      &   1.1$\pm$0.3 \\
S$_{2700}$(Jy)    &      &   3.5$\pm$1.0 \\ 
S$_{1420}$(Jy)    &      &   3.4$\pm$0.9 \\
S$_{408}$(Jy)     &      &   1.8$\pm$0.9 \\  
Spectral index, $\alpha$ &    &  \\
408 and 1420 MHz  &           &   +0.5$\pm$0.3 \\
1420 and 2700 MHz  &      &  +0.03$\pm$0.1  \\ 
Angular radius ($\arcmin)$ &         &  12$\pm$3 \\
Linear radius $R$(pc)          &         &  15$\pm$4   \\
$n_e$($f$=1)             &      &   4$\pm$1    \\
$M_{\rm ion}$($f$=1)(\msun)         &      &    3700$\pm$600     \\
$n_e$($f$=0.1-0.2)(cm$^{-3}$) &      &   15-9   \\ 
$M_{\rm ion}$($f$=0.1-0.2)(\msun)&      &    1100-1500  \\
$U_{\rm rad}$ (pc cm$^{-2}$)&    &    45     \\
$U_*$ (pc cm$^{-2}$)&      &  $\sim$128    \\
$EM$(pc cm$^{-6}$)  &      &  (2.4$\pm$1.7)$\times$10$^3$     \\

\hline
      \multicolumn{3}{c}{{\it IR}}   \\
\hline
S$_{60}$ (Jy)    &      &   $\sim$6.9$\times$10$^3$ \\ 
S$_{100}$ (Jy)    &      &  $\sim$2.25$\times$10$^4$ \\
Dust color temperature (K)  &       &   28-31 \\
Dust mass (\msun)   &            &  90$\pm$55 \\

\hline

     \multicolumn{3}{c}{$^{12}$CO}(1-0)   \\
\hline     
$(l,b)$ centroid of IB &   &  111$^{\circ}$5\arcmin,--0$^{\circ}$4\arcmin \\ 

Velocity range $\Delta v$(\kms)    &            &   --58.4 to --43.5 \\
Expansion velocity (\kms)    &            &   8$\pm$1 \\
Angular radius of the shell $R_{\rm mol}$ ($\arcmin)$ &         &  17.5$\pm$3.0 \\
Linear radius of the shell (pc)          &         &  18$\pm$5   \\
$H_2$ mean column density (cm$^{-2}$)       &          &  (2.9$\pm$1.3)$\times$10$^{20}$   \\
$H_2$ mass of the shell   (\msun)   &                 & (8.2$\pm$1.7)$\times$10$^2$             \\
\hline

\end{tabular}
\end{minipage}
\end{table}

\subsection{Origin and scenario}

     Assuming the CO expansion velocity  of 8$\pm$1 \kms\ listed in Table~\ref{tabla1}, the dynamical age of SG\,13, according to wind bubble evolutionary models, is $t_{\rm d} = 0.55 R_{\rm mol}/v_{\rm exp}= (1.4\pm0.4)$\x10$^6$ yr. Considering the radio continuum, IR and molecular counterpart of SG\,13, the kinematic energy of the interstellar bubble is  $E_k$=(9.8$\pm$6.0)$\times$10$^{47}$ erg.\

    Assuming typical values for the stellar wind of a WN\,5 star, a mass loss rate $\dot{M}$ = 1.75$\times$10$^{-5}$ \My\ and a terminal velocity $v_\infty$ = 1520 \kms\ (Smith et al. 2002, Cappa et al. 2004), and assuming a previous O-type phase with values of $\dot{M}$ = 2.3$\times$10$^{-7}$ \My\ and $v_\infty$ = 1950 \kms\ (Smith et al. 2002), we obtain a mechanical luminosity for the stellar wind of WR\,157 during the O and WR phases,  $L_O \simeq$3.5$\times$10$^{35}$ erg/s and $L_W \simeq$1.3$\times$10$^{37}$ erg s$^{-1}$, respectively. Bearing in mind a lifetime $t_O$ = 5$\times$10$^6$ yr for the O-type phase of a star with an initial mass of 40 \msun, and $t_{WR}$ = 0.4$\times$10$^6$ yr for the WR phase (Meynet et al. 1994), the mechanical energy injected into the interstellar medium in each evolutionary phase is $E_{\rm w_O} \simeq$ 5.5$\times$10$^{49}$ erg and $E_{\rm w_{WR}} \simeq$ 1.6$\times$10$^{50}$ erg.\

     The ratio $\epsilon=E_{\rm k}/E_{\rm w}$ estimated considering only the WR phase of WR\,157 is 6\x\ 10$^{-3}$. We can conclude that the stellar wind of the WR star alone is the main one responsible for shaping the interstellar bubble around the open cluster.\   

     The distributions of the interstellar dust and the ionized and molecular material in the environs of Mrk\,50 can be explained as the consequence of the action of the stellar winds of WR\,157. As regards \hi\ gas, no neutral atomic counterpart of SG\,13 could be identified from the present study, since the spatial distribution of  Shells A and B is very different from that of SG\,13. It is likely that the small amount of \hi\ gas resulting from the photodissociation of the molecular gas would be hardly detectable against the strong background \hi\ emission.

   As regards the ionized gas, we can suggest two possible scenarios:\

\begin{itemize}

\item The inner bright optical filaments were probably generated by the action of the present WR phase of the star. The shock fronts have ionized and dissociated the molecular circumstellar environment creating photodissociated regions. The outer optical filaments may have been swept up by the stellar winds in earlier stellar phases. 

\item The other possibility is that the distribution of the different optical filaments is far from being coplanar. If this were the case, we would be observing the projection of the filaments over the plane of the sky, the real dimensions of the structures being then larger than indicated in Table~\ref{tabla1}.\       
     
\end{itemize}

\section{Conclusion}

    Using the CGPS high-resolution radio continuum and 21 cm \hi\ line data, supplemented by previous optical, IR and CO surveys, we arrive at the following conclusions concerning the ring nebula SG13 and the associated star WR 157.\
 
The radio continuum emission correlates extremely well with the optical DSS R and IR 60 $\mu$m images. Interference effects from nearby Cas A limit our ability to reliably determine the spectral index between the two CGPS radio frequencies, however the spectral index between 1420 MHz (CGPS) and 2700 MHz (F$\ddot{u}$rst et al. 1990) is consistent with thermal emission.\

   A partial ring of CO emission in the range --56 to --43 \kms\ is seen to circumscribe the optical, IR (60 $\mu$m) and radio continuum emission of SG13. This range of velocities is consistent with previously measured velocity determinations by Lozinskaya et al. (1986) and Pedlar (1980), based on optical lines and the H166 radio recombination line, respectively.\

  An \hbox{analysis} of the spatial distribution of IR point sources having colours characteristic of YSOs shows that an excess of such sources appears projected over the molecular ring surrounding SG13, suggesting that star formation triggered by the stellar members of Mrk 50, in particular WR 157, is taking place.\

  The kinematics and dynamics of the gas, dust and molecular material around SG13 are entirely compatible with the hypothesis that the WR star alone is responsible for shaping the ISM around the open cluster Mrk 50.\

 A 50$\arcmin$ diameter \hi\ shell (shell A) is detected in the --32 to --50 \kms\ range to the north of SG13. Although this range partially overlaps the velocity range of the CO ring surrounding SG13, this \hi\ structure is entirely distinct from the other structures (molecular ring, optical, IR and radio continuum). The star WR157 appears projected onto the southern boundary of this \hi\ shell. An expansion velocity of 13$\pm$2 \kms\ is inferred from a velocity-position diagram. A second, smaller, \hi\ cavity (shell B) is also detected in the velocity range --52 to --41 \kms, suggesting a lower expansion velocity of 9$\pm$2 \kms. It too does not coincide with any of the structures associated with SG13. We conclude that these two cavities or shells are unlikely to be physically related to SG13.\

  The lack of an \hi\ shell, which has been detected associated with a large member of interstellar bubbles, is probably due to a low column density of the \hi\ gas resulting from the photodissociation of the molecular gas.\

\section{Acknowledgements}

  We acknowledge the referee, Dr. Peter Phillips, for his useful suggestions and comments. This project was partially financed by the Consejo Nacional de 
Investigaciones Cient\'{\i}ficas y T\'ecnicas (CONICET) of Argentina under 
project PIP 5886/05,  Universidad Nacional de La Plata (UNLP) under project 11/G072, and Agencia Nacional de Promoci\'on Cient\'{\i}fica y Tecnol\'ogica (ANPCYT) under project PICT 14018/03.
The Digitized Sky Survey (DSS) was produced at the Space Telescope Science
Institute under US Government grant NAGW-2166. This work was partly (S.P.) supported by the Natural Sciences
and Engineering Research Council of Canada (NSERC) and the Fonds
FQRNT of Qu\'ebec.  The  DRAO Synthesis
Telescope is operated as a national facility by the National Research 
Council of Canada. The CGPS is a Canadian project with
international partners and is supported by grants from NSERC.
Data from the CGPS
is publicly available through the facilities of the Canadian
Astronomy Data Centre (http://cadc.hia.nrc.ca) operated by the
Herzberg Institute of Astrophysics, NRC.

\end{document}